\documentclass[epj]{svjourAdelaide}
\usepackage{rotate}
\usepackage{epsfig}
\def\bge{\begin{equation}}
\def\ene{\end{equation}}
\newcommand{\nc}{\newcommand}
\nc{\non}{\nonumber}
\def\be{\begin{equation}}
\def\ee{\end{equation}}
\def\bga{\begin{eqnarray}}
\def\ena{\end{eqnarray}}
\def\eea{\end{eqnarray}}
\def\bg{\begin{eqnarray}}
\def\en{\end{eqnarray}}

\def\ra{\rightarrow}

\def\hbar{\not\!h}
\nc{\mpi}{m_\pi}
\nc{\cpt}{$\chi$pt}
\nc{\gev}{\ {\rm GeV}}
\nc{\mev}{\ {\rm MeV}}
\begin{document}
%
%
\title{Recent Developments in Quark Nuclear Physics}
\author{Anthony~W.~Thomas, Derek~B.~Leinweber \and Ross~D.~Young
}                     
%
%
\institute{Special Research Centre for the Subatomic Structure of
Matter and\\ Department of Physics and Mathematical Physics, University
of Adelaide, Adelaide 5005, Australia 
}
\date{Received: 30 September 2002 / Published online: 22 October 2003}
%
\abstract{ We provide an overview of recent work exploring the
quark-mass dependence of hadronic observables and the associated role
of chiral nonanalytic behavior due to the meson-cloud of hadrons.  In
particular, we address an issue of great current interest, namely the
degree of model independence of results obtained through a controlled
extrapolation of lattice QCD simulation results.  Physical insights
gained from this research are highlighted.  We emphasize how chiral
effective field theory formulated with a finite-range regulator
provides a reliable and {\it model-independent} extrapolation to the
physical world.
} 
\maketitle
\section{Introduction}
 
Quark nuclear physics describes our attempts to understand the structure
of hadronic systems, including nuclei and dense matter, in terms of  
quarks and gluons -- the fundamental degrees of freedom in QCD. It is
impossible in just a few pages to provide even a vague outline of the
many exciting physics issues currently being addressed in this field --
{}from the possible phase transition to one or more quark-gluon phases
at high temperature or density \cite{McLerran:2002jb}, to   
suggestions of changes of hadron properties in-medium
\cite{Dieterich:2000mu,Lu:1998tn}. Instead we shall concentrate on just
one development which offers considerable insight into hadron structure
from QCD itself, an approach which has led to surprisingly accurate
comparisons between lattice QCD data and experiment as well as
remarkable insights into how one might improve hadron models.

As the time for calculations within lattice QCD \cite{Rothe:kp} with 
dynamical fermions  (including
$q-\bar q$ creation and annihilation in the vacuum) scales 
as $m_q^{-3.6}$ \cite{Lippert:zq},  
current calculations have been limited to light quark
masses 6--10 times larger than the physical ones.
With the next generation of
supercomputers, around 10 Teraflops, it should be possible
to get as low as 2--3 times the physical quark mass, but to actually reach
that goal on an acceptable volume will require at least 500
Teraflops.  This is probably 10-20 years away.

Since a major motivation for lattice QCD must be to unambiguously
compare the calculations of hadron properties with experiment, this is
somewhat disappointing.  The only remedy for the next decade at least is
to find a way to extrapolate masses, form-factors, and so on, calculated
at a range of masses considerably larger than the physical ones, to the
chiral limit.  In an effort to avoid theoretical bias this has usually
been done through low-order polynomial fits as a function of quark
mass. Unfortunately, as we discuss in Sec.~2, this is incorrect 
and can yield quite misleading
results because of the Goldstone nature of the pion.

The essential problem in performing calculations at realistic quark
masses (of order 5 MeV) is the approximate chiral symmetry of QCD.
Goldstone's theorem tells us that chiral symmetry is dynamically
broken and that the non-perturbative vacuum is highly non-trivial
\cite{Thomas:2001kw}, with massless Goldstone bosons in the limit $m_q
\rightarrow 0$.  For finite quark mass these bosons are the three
charge states of the pion with a mass $m_\pi^2 \propto m_q$.  Although
this result strictly holds only for $m_\pi^2$ near zero (the
Gell-Mann--Oakes--Renner relation), lattice simulations show it is a
good approximation for $m_\pi^2$ up to 1 GeV$^2$ or so, and we shall
use $m_\pi^2$ here as a measure of the deviation from the chiral
limit.

On these very general grounds, one is therefore compelled to
incorporate the non-analyticity into any extrapolation procedure.  The
classical approach to this problem is chiral perturbation theory
($\chi$pt), an effective field theory built upon the symmetries of QCD
\cite{Gasser:1983yg}.  There is considerable evidence that the scale
naturally associated with chiral symmetry breaking in QCD,
$\Lambda_{\chi {\rm SB}}$, is of order $4 \pi f_\pi$, or about 1 GeV.
$\chi$pt then leads to an expansion in powers of $m_\pi/\Lambda_{\chi
{\rm SB}}$ and $p/\Lambda_{\chi {\rm SB}}$, with $p$ a typical
momentum scale for the process under consideration.  At ${\cal
O}(p^4)$, the corresponding effective Lagrangian has only a small
number of unknown coefficients which can be determined from
experiment.  On the other hand, at ${\cal O}(p^6)$ there are more than
100 unknown parameters \cite{Fearing:1994ga}, far too many to
determine phenomenologically.

While this situation seems formidable, the resolution is already in
hand. We must first realize that the lattice data obtained so far
represents a wealth of information on the properties of hadrons within
QCD itself over a range of quark masses. 
Just as the study of QCD as a function of
$N_c$ has taught us a great deal, so the behaviour as a function of
$m_q$ can yield considerable insight into hadronic physics.

The first thing that stands out, once one views the data as a whole,
is just how smoothly every hadron property behaves in the region of
large quark mass.  In fact, baryon masses behave like $a + b\, m_q$,
magnetic moments like $(c + d\, m_q)^{-1}$, charge radii squared like
$(e + f\, m_q)^{-1}$ and so on. Thus, if one defined a light
``constituent quark mass'' as $M \equiv M_0 + \tilde{c}\, m_q$ (with
$\tilde{c} \sim 1$), one would find baryon masses proportional to $M$
(times the number of u and d quarks), magnetic moments proportional to
$M^{-1}$ and so on -- just as in the constituent quark picture.  There
is little or no evidence for the rapid, non-linearity associated with
the branch cuts created by Goldstone boson loops.

Over the past few years we have come to a deep understanding of why
QCD exhibits these features. It will be helpful to summarise those
conclusions here:
\begin{itemize}
\item In the region of quark masses $m_q > 60$ MeV or so ($m_\pi$
greater than typically 400-500 MeV)
hadron properties are smooth, slowly varying
functions of something like a constituent quark mass, $M \sim M_0 + c\,
m_q$ (with $c \sim 1$).
\item Indeed, $M_N \sim 3\, M, M_{\rho, \omega} \sim
2\, M$ and magnetic moments behave like $1/M$.
\item As $m_q$ decreases below 60 MeV or so, chiral symmetry leads to
rapid, non-analytic variation, with $\delta M_N \sim {m_q}^{3/2},
\delta \mu_H \sim {m_q}^{1/2}$ and \\ $\delta <r^2>_{\rm ch} \sim \ln
m_q$.
\item Chiral quark models like the cloudy bag
\cite{Thomas:1982kv,Miller:1979kg,Theberge:1981mq} provide a natural
explanation of this transition. The scale is basically set by the
inverse size of the of the composite source, above which chiral loops
are strongly suppressed. Below this scale the pion Compton wavelength
is larger than the source and one begins to see rapid, non-analytic
chiral corrections.
\end{itemize}

These are remarkable results that will have profound consequences for
our further exploration of hadron structure within QCD as well as the
analysis of the vast amount of data now being taken concerning unstable
resonances. In terms of immediate results for the structure of the
nucleon, we note that the careful incorporation of the correct chiral
behaviour of QCD into the extrapolation of its properties calculated on
the lattice has produced:
\begin{enumerate}
\item The most accurate values of the proton and neutron 
magnetic moments from lattice QCD \cite{Leinweber:1999ej}.
\item The most accurate value of the sigma 
commutator from lattice QCD \cite{Leinweber:2000sa}.
\item The most accurate values for the charge radii of the baryon
octet from lattice QCD \cite{Hackett-Jones:2000js}.
\item The most accurate values for the magnetic moments of the
hyperons from lattice QCD \cite{Hackett-Jones:2000qk,Hemmert:2002uh}.
\item Good agreement between the extrapolated moments of the
non-singlet distribution, $u - d$, calculated in lattice QCD and the
experimentally measured moments \cite{Detmold:2001jb,Detmold:2001dv}.
\item The most accurate estimates of the low moments of the
spin-dependent parton distribution functions at the physical quark
mass from lattice QCD \cite{Detmold:2002nf}.
\item An understanding of the failures of the assumption of
universality of quark electromagnetic properties
\cite{Leinweber:2001ui} and an improved lattice estimate of the
strangeness magnetic moment of the proton $G_M^s$
\cite{Leinweber:1999nf}.
\end{enumerate}
{}Furthermore, this approach, together with the observed
constituent-quark-like behaviour seen in the lattice data for $m_q > 50$
MeV (as noted earlier), has suggested a novel way of modelling hadron
structure \cite{Cloet:2002eg,Cloet2}.

Apart from the original publications, these 
developments have been fairly widely reported at various
conferences -- e.g. see Refs.~\cite{Detmold:2001hq,Thomas:2001iv}.
Here we focus particularly on the question of the
extrapolation of hadron masses in order to clarify an issue of great
current interest, namely the degree of model independence of the results
obtained after a controlled chiral extrapolation.

\section{Effective Field Theory}
Chiral perturbation theory is a low-energy effective field theory for
QCD. Low-energy properties of QCD can be expanded about the limit of
vanishing momenta and quark mass. In relation to the extrapolation of
lattice data, $\chi$pt provides a functional form applicable in the
limit $\mpi\to 0$. 

Goldstone boson loops give rise to specific corrections to baryon
properties --- most importantly, they give rise to non-analytic
behaviour as a function of quark mass. The low-order, non-analytic
contributions arise from the pole in the Goldstone boson propagator
and hence are {\em model-independent} \cite{Li:1971vr}. Analytic
variation of hadron properties is not constrained via the symmetry and
hence expansions contain free parameters which must be determined by
comparison with data.

Effective field theory then tells us that the general expansion of the
nucleon mass about the SU(2) chiral limit is:
\bga
m_N & = & \alpha_0 + \alpha_2\, m_\pi^2 + \alpha_4\, m_\pi^4 \non \\
    &   &  + \sigma_{{\rm N} \pi}(m_\pi, \Lambda) + \sigma_{\Delta \pi}(m_\pi, \Lambda)+\ldots\, ,
\label{eq:eftexp}
\eea
where $\sigma_{{\rm B} \pi}$ is the self-energy arising from a $B\pi$
loop ($B$ = $N$ or $\Delta$). The expansion has been written
explicitly in this form to highlight that the theory is equivalently
defined for an arbitrary regulator --- see Ref.~\cite{Donoghue:1998bs}
for a complete discussion.

The traditional approach within the literature is to use dimensional
regularisation to evaluate the self-energy integrals. Under such a
scheme the $NN\pi$ contribution simply becomes $\sigma_{{\rm N}
\pi}(\mpi,\Lambda)\to c_{\rm LNA} \, \mpi^3$ and the analytic terms,
$\alpha_n\, \mpi^n$ undergo an infinite renormalisation. The $\Delta$
contribution behaves similarly, producing a logarithm and one obtains
a series expansion of the nucleon mass about $\mpi= 0$
\bga 
m_N &=& c_0 + c_2\, m_\pi^2 + c_4\, m_\pi^4 \non \\
    & & + c_{\rm LNA}\, m_\pi^3 + c_{\rm NLNA}\, m_\pi^4 \ln m_\pi + \ldots\, ,
\label{eq:drexp}
\ena
where the $\alpha_i$ have been replaced by the renormalised (and finite)
parameters $c_i$.

It is not clear, a priori, that any such truncated expansion will be
capable of reliably fitting lattice data. The first empirical
indication of serious problems in this approach came with the
realization that lattice data could not recover the 
{\em model-independent} coefficient, $c_{\rm LNA}$. Truncating the power
series at the $\mpi^3$ term and allowing $c_{\rm LNA}$ to vary as a
free fit parameter, together with $c_0$ and $c_2$, produced a value
$c_{\rm LNA} \sim -0.76\gev^{-2}$ \cite{Leinweber:1999ig}. This should
be compared with the physical value of $-5.6\gev^{-2}$ --- a factor of
8 larger! This tells us immediately that {\bf either} there are
serious convergence problems with the third order expansion {\bf or}
lattice QCD is in error.  Clearly most readers would opt for the first
possibility and so do we.

Even by retaining all terms as described in Eq.~(\ref{eq:drexp}) it is
not clear that reliable extrapolation can be guaranteed by fitting
lattice data over a range of (heavy) quark masses.  One point of issue
is that it is derived in the limit $m_\pi \ll \Delta (\equiv m_\Delta
- m_N)$, whereas the lowest lattice data with dynamical fermions that
one can expect in the next decade is perhaps 200-250 MeV --
c.f. $\Delta = 292$ MeV. It should be clear to those familiar with
lattice simulations that even at this lightest pion mass the branch
cut will not be observed due to the restricted phase space on a finite
volume lattice.  Consequently all lattice data will still lie above
$\Delta$. Mathematically the region $m_\pi \approx \Delta$ is
dominated by a square root branch cut which starts at $m_\pi =
\Delta$. Using dimensional regularisation this takes the form
\cite{Banerjee:1995wz}:
\bga
&& \frac{6 g_A^2}{25 \pi^2 f_\pi^2}\left[ (\Delta^2 - m_\pi^2)^{\frac{3}{2}}
\ln (\Delta + m_\pi -\sqrt{\Delta^2 - m_\pi^2}) \right. \non \\
&& \left. - \frac{\Delta}{2}
(2 \Delta^2 - 3 m_\pi^2) \ln m_\pi \right],
\label{eq:log}
\ena
for $m_\pi < \Delta$, while for $m_\pi > \Delta$ the first logarithm
becomes an arctangent. No serious attempt has been made to extend the
formal expansion in Eq.~(\ref{eq:drexp}) to incorporate this cut in an
analysis of lattice data and, given the number of parameters to be
determined if one works to order $m_\pi^6$, it is not likely that it
will be done in the next decade.

Even ignoring the $\Delta \pi$ cut for a short time, studies of the
formal expansion of the N~$\ra $N~$\pi \ra $~N
self-energy integral, $\sigma_{{\rm N} \pi}$, suggest that it has
abysmal convergence properties. Using a sharp, ultra-violet cut-off,
Wright showed \cite{SVW} that the series diverged for $m_\pi > 0.4$ GeV.
If one instead uses a dipole cut-off, which in view of the
phenomenological shape of the nucleon's axial form-factor is much more
realistic, it is worse -- with the radius of convergence being around
0.25 GeV. We return to this in Sect.~3.

The main issue of the convergence of this truncated series,
Eq.~(\ref{eq:drexp}), has its origin fixed in the formalism that it is
derived from the general form of Eq.~(\ref{eq:eftexp}). The
dimensionally regulated approach requires that the pion mass remain
much lighter than every other mass scale involved in the problem. This
requires that $\mpi/\Lambda_{\chi {\rm SB}}\ll 1$ and $\mpi/\Delta\ll
1$. A further scale, as addressed in the introduction, is set by the
physical extent of the source of the pion field. This scale,
$\Lambda\sim R_{SOURCE}^{-1}$, corresponds to the transition between
rapid, non-linear variation and smooth, {\em constituent-like} quark
mass behaviour. An alternative procedure would be to regulate
Eq.~(\ref{eq:eftexp}) with a finite $\Lambda$ which physically
corresponds to the source of the meson cloud having an extended
structure.

In summary, the low-energy effective field theory can be very useful
in describing the quark mass behaviour of hadron properties. These
powers have unfortunately been lost in the literature, where only a
single type of regulator (i.e. dimensional) has been studied in detail.

\section{Accurate, Model-Independent Method of Chiral Extrapolation}
We now turn to the direct application of Eq.~(\ref{eq:eftexp}),
written in a regulator independent form, to the extrapolation of
lattice data. By studying a range of different regulators, both finite
ranged and the dimensional approach, we can study the sensitivity to
the form chosen.

The additional analytic term, $\alpha_4\, \mpi^4$, differs from previous
studies using a finite ranged regulator \cite{Leinweber:1999ig}.
Since one is working to non-analytic order $\mpi^4\log\mpi$ one is
certainly permitted freedom in $\alpha_4$ \cite{Donoghue:1998bs}.  In
practice, whether or not this can be reliably determined, together
with $\Lambda$, will depend on the data available.

The self-energies, $\sigma_{{\rm B} \pi}$, are evaluated with a
variety of regulators. Included here are both that of the truncated
power series resulting from dimensional regularisation and also those
of finite range, including sharp cut-off, monopole, dipole and
Gaussian. The coefficients of non-analytic terms are constrained to
their phenomenological values, while the parameters $\alpha_i$ and
$\Lambda$ are then determined by fitting {\em lattice data}.

It is again worth noting that, independent of regulator, precisely the
same expansion of Eq.~\ref{eq:drexp} will be obtained in the limit
$\mpi\ll \Lambda$. Although the parameters $\alpha_i$ {\bf will}
depend upon the choice of regulator, one can expand the self-energy
terms to order $m_\pi^2$ (or higher) to obtain the chiral coefficients
at the appropriate order and compare with the truncated expansion of
Eq.~(\ref{eq:drexp}) --- see Ref.~\cite{Donoghue:1998bs} for a full
discussion of this issue.

\begin{figure}[tb]
\begin{center}
\includegraphics[angle=0,width=\columnwidth]{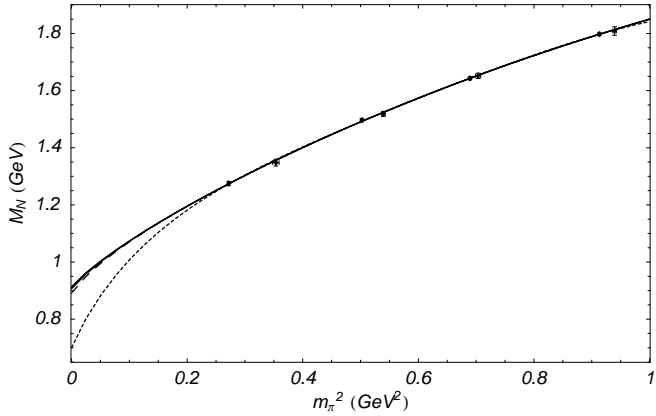}
\caption{Fits to lattice data \protect\cite{AliKhan:2001tx}
for five different ultra-violet regulators.}
\label{fig:latfit}
\end{center}
\end{figure}
The results of our fits to lattice data studied with a range of
regulators are shown in Fig.~\ref{fig:latfit}. The short-dash curve
shows the fit obtained using the dimensionally regulated form
of Eq.~\ref{eq:drexp} extended to a further term in the expansion,
$\alpha_6\, \mpi^6$.  This additional term is necessary for a more
reasonable fit due to the large NLNA contribution. The long dash curve
is the result of using a sharp cut-off regulator, while the following
three fits (solid curves), {\bf indistinguishable on this plot}
correspond to the monopole, dipole and Gaussian regulated forms.

\begin{figure}[tb]
\begin{center}
\includegraphics[angle=0,width=\columnwidth]{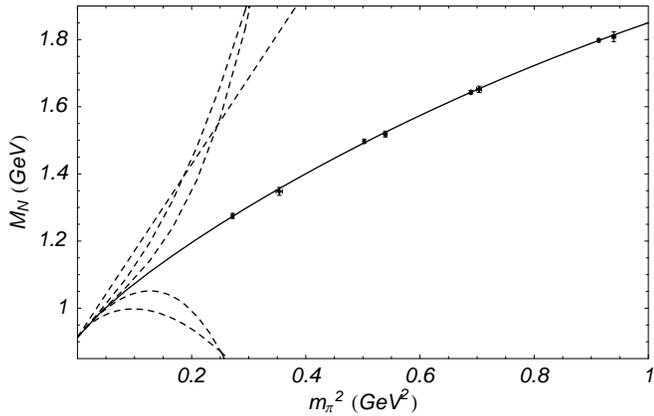}
\caption{The fit to the lattice data using the dipole regulator.  The
dashed curves show power series expansions of this fit to successive
orders in $m_\pi$ for $\mpi^2\to\mpi^6$.}
\label{fig:dipexp}
\end{center}
\end{figure}

The extrapolation of lattice data is clearly independent of the choice
of finite, ultra-violet regulator. Knowing this we can examine the
range of convergence of a truncated power series. Selecting the dipole
form and neglecting the $\Delta\pi$ contribution to the nucleon self
energy one can obtain a closed analytic expression. An expansion in
powers of the pion mass is shown in Fig.~\ref{fig:dipexp}. Here we see
that convergence of the expansion to order $\mpi^6$ breaks down above
$\mpi\sim 250\mev$.

For the reasons outlined, it is essential that the self-energies are
evaluated using some ultra-violet regulator --- a sharp cut-off or a
dipole form, for example. Whatever is chosen does not effect the
non-analytic structure which is guaranteed correct. The branch points
at $m_\pi$ equals zero and $\Delta$ are incorporated naturally. The
use of a finite regulator then automatically produces the transition
scale associated with the physical extent of the meson source.

The essential point is that studies of the nucleon
(c.f. Ref.~\cite{SVW} and Fig.~\ref{fig:latfit}), the $\Delta$
(c.f. Fig.~4 of Leinweber {\it et al.} \cite{Leinweber:1999ig}) and
the $\rho$ meson \cite{Leinweber:2001ac} suggest that {\it this
procedure will result in little or no model dependence in the
extrapolation to the physical pion mass once there is accurate lattice
data for $m_\pi \sim 0.3\gev$ or less}. Physically this is possible
because the self-energy loops are rapidly suppressed in the region
$m_\pi > 0.4\gev$. Thus, an extrapolation based on
Eq.~(\ref{eq:eftexp}) formulated with the selection of a long-distance
regulator allows one to respect {\it all the chiral constraints}, keep
the number of fitting parameters low and yield essentially
model-independent results at the physical pion mass. No other approach
can do this.

\section{Possible Connection to QQCD}
Although quenched QCD (QQCD) is an unphysical theory, it provides an
alternative avenue for enhancing our understanding of chiral
extrapolation. Multi-mass techniques allow a dense set of quark masses
to be simulated with relative computational ease. Together with recent
advances in numerical techniques, which allow simulations to be
performed at light quarks masses \cite{Zanotti:2001yb}, one will be
able to very accurately determine the quark mass dependence of
quenched simulations within the light quark mass regime.

The study of baryon spectroscopy in quenched lattice QCD has recently
made great progress. We have already noted that the lattice data
behaves like a constituent quark model for quark masses above 50--60
MeV because Goldstone boson loops are strongly suppressed in this
region. This not only provides a very natural explanation of the
similarity of quenched and full data in this region but it also
suggests a much more ambitious approach to hadron spectra. It suggests
that one might remove the small effects of Goldstone boson loops in
QQCD (including the $\eta'$) and then estimate the hadron masses in
full QCD by introducing the Goldstone loops which yield the LNA and
NLNA behaviour in full QCD.

\begin{figure}[htb]
\begin{center}
\includegraphics[angle=90,width=\columnwidth]{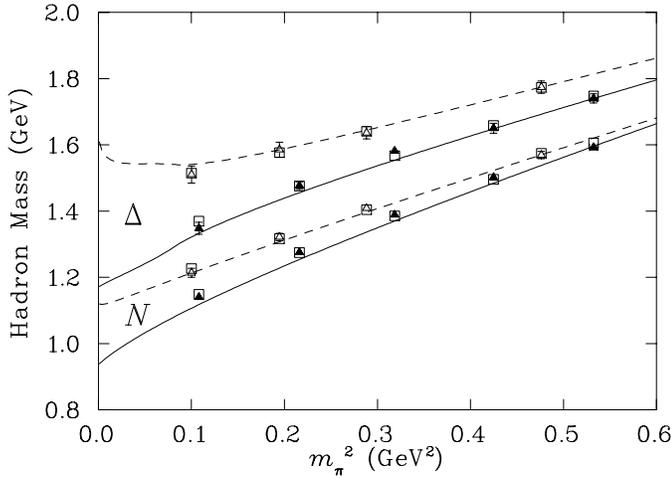}
\caption{Fits to both quenched (open triangles) and dynamical (closed
triangles) lattice data \protect\cite{Bernard:2001av} using a dipole
regulator \protect\cite{Young:2002cj}.  }
\label{fig:qfit}
\end{center}
\end{figure}

As a first test of this idea, Young {\it et al.} \cite{Young:2002cj}
recently analysed the MILC data \cite{Bernard:2001av} for the N and
$\Delta$, using Eq.(\ref{eq:eftexp}) (with $\alpha_4=0$) for full QCD
and the appropriate generalization for QQCD -- i.e. using quenched
pion couplings as well as the single- and double-``hairpin'' $\eta'$
loops \cite{Labrenz:1996jy,Young:rx}. The results illustrated in
Fig.~\ref{fig:qfit} are remarkable.  The values of $\alpha_0$ and
$\alpha_2$ for the N (or the $\Delta$) obtained in QQCD agree
within statistical errors with those obtained in full QCD.  Certainly
this result (unlike the result for the extrapolation of individual
hadron masses as noted above) is somewhat dependent on the shape of
the ultra-violet cut-off chosen -- although the extent of that is yet
to be studied in detail.  Nevertheless, given that the study involved
the phenomenologically favoured dipole form, it is a remarkable result
and merits further investigation.

\section{Conclusion}

At the present time we have a wonderful conjunction of opportunities.
Modern accelerator facilities are providing data of unprecedented
precision over a tremendous kinematic range at the same time as
numerical simulations of lattice QCD are delivering results of
impressive accuracy.  It is therefore timely to ask how to use these
advances to develop a new and deeper understanding of hadron structure
and dynamics.

We have demonstrated that the use of chiral effective field theory can
provide accurate extrapolation formulae. In particular, we have shown
that the extrapolation of the nucleon mass exhibits minimal
model-dependence in the choice of finite-ranged, ultra-violet
regulator.

In combination with the very successful techniques for chiral
extrapolation, lattice QCD will finally yield accurate data on the
consequences of non-perturbative QCD.  Furthermore, the physical
insights obtained from the study of hadron properties as a function of
quark mass will guide the development of new quark models and hence a
much more realistic picture of hadron structure.

\section*{Acknowledgements}
We would like to thank those colleagues who have contributed to our
understanding of the problems discussed here, notably Jonathon Ashley,
Ian Cloet, Will Detmold, Wally Melnitchouk, Stewart Wright and James
Zanotti. This work was supported by the Australian Research Council
and the University of Adelaide.

\end{document}